\documentclass[proof]{pasj00}
\Received{2015 July 15}
\Accepted{2015 October 17}
\SetRunningHead{Suzaku Observation of G272.2$-$3.2}{Kamitsukasa et al.}

\usepackage{times}
\usepackage{lscape}
\usepackage[dvips]{graphicx}
\usepackage{bm}
\renewcommand{\arraystretch}{1}

\begin{document}

\title{Suzaku study on the Ejecta of the Supernova Remnant G272.2$-$3.2}
\author{Fumiyoshi \textsc{Kamitsukasa},\altaffilmark{1,2}
        Katsuji \textsc{Koyama},\altaffilmark{1,2,3}
        Hiroshi \textsc{Nakajima},\altaffilmark{1,2}
        Kiyoshi \textsc{Hayashida},\altaffilmark{1,2}
        Koji \textsc{Mori},\altaffilmark{4}
        Satoru \textsc{Katsuda},\altaffilmark{5}
        Hiroyuki \textsc{Uchida} \altaffilmark{3}
        and Hiroshi \textsc{Tsunemi},\altaffilmark{1,2}}
\altaffiltext{1}{%
   Department of Earth and Space Science, Osaka University, 1-1 Machikaneyama-cho,
   Toyonaka, Osaka 560-0043, Japan}
\altaffiltext{2}{%
   Project Research Center for Fundamental Sciences, Osaka University, 1-1 Machikaneyama-cho,
   Toyonaka, Osaka 560-0043, Japan}
\altaffiltext{3}{%
   Department of Physics, Graduate School of Science, Kyoto University, 
   Kitashirakawa Oiwake-cho, Sakyo-ku, Kyoto 606-8502, Japan}
\altaffiltext{4}{%
   Department of Applied Physics and Electronic Engineering, Faculty of Engineering, 
   University of Miyazaki, \\1-1 Gakuen Kibanadai-Nishi, Miyazaki 889-2192, Japan}
\altaffiltext{5}{%
   Institute of Space and Astronautical Science, 
   3-1-1, Yoshinodai, Sagamihara, Kanagawa 229-8510, Japan}

\email{kamitsukasa@ess.sci.osaka-u.ac.jp}


\maketitle

\begin{abstract}

We report re-analyses of the Suzaku observations of the Galactic supernova remnant (SNR), G272.2$-$3.2, for which the previous studies were limited  below 3\,keV. With careful data reduction and background subtraction, we discover the K-shell lines of Ar, Ca, and Fe above 3\,keV.  The X-ray spectrum of G272.2$-$3.2 consists of two components, a low-temperature collisional ionization equilibrium (CIE) plasma ($kT_{\rm e} \sim 0.2$\,keV) and a high-temperature non-equilibrium ionization (NEI) plasma ($kT_{\rm e} = 0.6$--$3$\,keV). The CIE plasma has solar abundances over the entire area, hence it would originate from the interstellar medium. On the other hand, the abundances of the NEI plasma increase toward the inner region, suggesting the ejecta origin.  The line center energy of the Fe K-shell emission ($\sim 6.4$\,keV) suggests that the ejecta are recently heated by the reverse shock, a common feature in Type Ia SNRs.

\end{abstract}

\section{Introduction}

Supernovae (SNe) originate from two types of progenitors. One is an explosive runaway of thermal nuclear fusion (Type Ia SN), and the other is a gravitational core-collapse of massive stars (CC SN). For many supernova remnants (SNRs), it is not evident which type of a progenitor is a true origin. The key of the typing is the measurement of SN ejecta. A Type Ia SN produces a large number of intermediate mass (Si--Ca) and Fe-peaked elements, while the major products of a CC SN are light elements (He--Mg). Since an interstellar medium (ISM) is also taken into remnants, it is essential to accurately separate ejecta from an ISM to distinguish between SN types.

G272.2$-$3.2 is one of thermal dominated SNRs in X-rays. It was firstly discovered with the ROSAT all sky survey (\cite{Greiner1993a}; \cite{Greiner1994a}). The ROSAT observations revealed the spherical structure of $\timeform{7.6'}$ radius, showing center-filled thermal X-rays. \citet{Greiner1994a} argued that the origin of the center X-ray emission is reverse-shocked ejecta from the SN explosion or cloud evaporation in the interior of the remnant. 

More recently, \citet{Harrus2001a} found a non-equilibrium ionization (NEI) plasma of $kT_{\rm e} \sim 0.7\,{\rm keV}$ with the ASCA and the ROSAT data. The statistical analysis of the Galactic SNRs distribution led to a probable distance of $2\,{\rm kpc}$. However, a large upper limit of $10\,{\rm kpc}$ was also reported using the interstellar absorption. \citet{Lopez2011a} suggested a Type Ia origin for G272.2$-$3.2 with a morphological analysis of the Chandra data.  In the Suzaku data, \citet{Sezer2012a} found an NEI plasma of super-solar abundances and a stratified structure. They claimed that the plasma originated from the ejecta of a Type Ia SN based on the abundance pattern. \citet{McEntaffer2013a}, using the Chandra data, confirmed the NEI plasma of $kT_{\rm e} = 0.7$--$1.5\,{\rm keV}$. They also found a cool collisional ionization equilibrium (CIE) plasma of $kT_{\rm e} \sim 0.2\,{\rm keV}$ in the outer regions. Since the abundances of the NEI plasma were depleted in the outer region, they claimed that the plasma in the center region was ejecta origin, while the outer region was the shock-heated ISM. \citet{Sanchez-Ayaso2013a}, using the XMM-Newton and the Chandra data, reported the two NEI plasmas similar to those by \citet{McEntaffer2013a}, but the temperature distribution was different; the temperatures in the center and outer plasmas were $0.76\pm0.03\,{\rm keV}$ and $1.05\pm0.79\,{\rm keV}$, respectively.

Based on the super solar abundances and abundance pattern of Si, S, and Fe, the authors in the previous studies proposed a Type Ia origin for G272.2$-$3.2 (table 1 of \cite{Sezer2012a}, table 5 of \cite{McEntaffer2013a}, and table 2 of \cite{Sanchez-Ayaso2013a}).  It is consistent with the fact that a pulsar or a pulsar wind nebula associated with the remnant was found neither in the X-ray (\cite{McEntaffer2013a}) nor the radio band (\cite{Duncan1997a}). However, most of the spectral analyses in the previous studies were limited below 3\,keV, and hence reliable K-shell line fluxes (or abundances) were available only for Ne, Mg, Si and S. In order to obtain more reliable spectra and abundances in the wider elemental range, we re-visit the Suzaku data, because the previous report by \citet{Sezer2012a} was poor in the data-reduction and analysis. They subtracted the background data from a very small region near the edge of the detector/mirror field, and hence degraded the S/N of the Suzaku spectra, especially above 3\,keV. We therefore make careful background estimation, data reduction and analysis.  As a result, we discover significant emission above 3\,keV for the first time, which includes K-shell lines of Ar, Ca and Fe. The high quality spectrum enables us to separate the ejecta from the ISM component and provides more reliable information for the typing of G272.2$-$3.2. This paper reports the SNR type and structure of G272.2$-$3.2 by these detailed study of abundances and morphology.

\section{Observation and Data Reduction}

G272.2$-$3.2 was observed with the X-ray Imaging Spectrometer (XIS; \cite{Koyama2007a}) on board the Suzaku satellite (\cite{Mitsuda2007a}) on May 28 (Obs ID: 506060010) and  November 12 (Obs ID: 506060020) in 2011. We employed three XISs on the focal planes of the X-ray Telescopes (XRTs; \cite{Serlemitsos2007a}). Two of the XISs are front-illuminated (FI) CCDs, while the other is a back-illuminated (BI) CCD. We used the version 6.16 of the HEAsoft tools for the data reduction and analysis. The archival data were reprocessed with the calibration data base (CALDB) released in 2014 February. The total exposure times of the first and the second observations, after the standard screening\footnote{http://heasarc.nasa.gov/docs/suzaku/processing/criteria\_xis.html}, were about 130\,ks and 26\,ks, respectively.

\section{Analyses and Results}

\subsection{Imaging Analysis}

Figure \ref{image} shows the narrow band images of the G272.2$-$3.2 in the energy bands of 0.86--0.96\,keV and 1.79--1.93\,keV, which include the K-shell lines of Ne and Si, respectively. We combine all the XIS data of the two observations to maximize the photon statistics. The emission of the Ne K-shell band extends over the whole region of the SNR, while that of Si is concentrated at the center. In the Ne-band image, a hot spot shown in a green ellipse is also found in the west, which is reported in the previous study (\cite{Harrus2001a}; \cite{McEntaffer2013a}; \cite{Sanchez-Ayaso2013a}). The narrow band image of Mg K-shell (1.28--1.42\,keV) has a similar profile to that of the Ne K-shell band, and those of S K-shell (2.41--2.51\,keV) and Fe L-shell (1.18--1.28\,keV) are similar to that of the Si K-shell band. 

\subsection{Spectral Analysis}

We employ the data of the first epoch observation in the following analysis. We use the XSPEC software version 12.8.2. The redistribution matrix files (RMFs) and the ancillary response files (ARFs) are generated with \texttt{xisrmfgen} and \texttt{xisarfgen} (\cite{Ishisaki2007a}), respectively. We estimate the non X-ray background (NXB) using \texttt{xisnxbgen} (\cite{Tawa2008a}). We adopt the solar abundances of \citet{Anders1989a}. All errors represent $1\sigma$ confidence levels.

\subsubsection{Background estimation}

Since G272.2$-$3.2 extends widely over the XIS field of view ($\timeform{17.8'}\times \timeform{17.8'}$), the available background region is limited in the region of $r > \timeform{510''}$  from the SNR center (here, BG region). Moreover, the BG region is contaminated by a leakage of the SNR emission due to the point spread function of the XRT with a relatively large half power diameter of $\sim \timeform{2'}$. To estimate the leaked SNR emission, we make spectra from the outer SNR region ($r = \timeform{300-450''}$) and the BG region. The NXBs are subtracted from the spectra of the two regions. The NXB-subtracted spectra consist of the SNR emission and the X-ray background (XBG).  The SNR emission is modeled by a bremsstrahlung continuum plus several Gaussian lines. The XBG consists of the cosmic X-ray background (CXB) and the Galactic halo (GH) (e.g., \cite{Kushino2002a}; \cite{Henley2013a}), which are represented by a power-law and a CIE (APEC in the XSPEC), respectively.  With these settings, we fit the two spectra simultaneously using the simulated ARFs for the two regions. The flux of the CXB is a free parameter within the range of the possible fluctuation ($\sim 40\,\%$) (\cite{Kushino2002a}). The fit is reasonable with $\chi^2_\nu$ (d.o.f.) = 1.19 (360). The best-fit parameters for the XBG is listed in table \ref{bkg-para}. We employ the XBG spectrum in the following SNR analysis.   

\subsubsection{Analysis of the whole SNR}

We first produce a spectrum from the whole region of the SNR (the circle of $r = \timeform{450''}$), and subtract the NXB. The NXB-subtracted spectrum is shown in figure \ref{whole_spec}.  Many K-shell lines from highly ionized Ne, Mg, Si, S, Ar, and Ca are found. The Ar and Ca lines are the first discovery from this SNR.  In addition, we find a line near at  $ 6.4\,{\rm keV}$.  With a Gaussian line fit, the center energy and flux are determined to be  $6.42_{-0.05}^{+0.07}\,{\rm keV}$ and $(1.83\pm0.50) \times 10^{-6}\,{\rm photons\,s^{-1}\,cm^{-2}}$, respectively. The center energy constrains the charge state of Fe to be ${\bf \le 18+}$ (\cite{Beiersdorfer1993a}).  This line is not a contamination of the Fe K-shell line in the Galactic ridge X-ray emission (GRXE), because the GRXE has the emission line at about 6.7 keV, and the flux of the GRXE near G272.2$-$3.2 should be very faint due to the large distance from the Galactic ridge (e.g., \cite{Uchiyama2013a}). 

We fit the whole region spectrum with an NEI plasma model (VNEI) adding the XBG model given in table 1. In the VNEI fit, the electron temperature $kT_{\rm e}$, column density $N_{\rm H}$  and ionization timescale $n_{\rm e}t$ are free parameters, where $n_{\rm e}$ and $t$ represent the electron density and the time after the shock heating, respectively.  The abundances of Ne, Mg, Si, S, Ar, and Fe are also free parameters, while those of Ca and Ni are tied to those of Ar and Fe, respectively. This fit, however, leaves a line-like residual around $1.2\,{\rm keV}$ that comes from the uncertainty of the Fe L-shell data in the VNEI code (e.g., \cite{Borkowski2006a}; \cite{Yamaguchi2011a}). Thus we add a Gaussian line at $1.2\,{\rm keV}$. The $\chi^2_\nu$ (d.o.f.) is largely improved from 1.89 (2006) to 1.38 (2004), but residuals below $1\,{\rm keV}$ are still large (panel:(a) in figure \ref{whole_spec}).  Therefore, we add a CIE model (APEC), where we leave the abundances free. Although this fit largely improved the $\chi^2_\nu$ (d.o.f.) from 1.38 (2004) to 1.21 (2001), this model fit is statistically rejected. 

Significant residuals are found at 6.4 keV and around 2--3 keV (panel:(b)).  The former is due to the K-shell line of nearly neutral Fe, and the latter is due to K-shell complexes of Si and S.  Thus, for further improved fit, we divide the VNEI spectrum into four components, they stand for H--Mg (VNEI\,1), Si--S (VNEI\,2), Ar--Ca (VNEI\,3), and Fe--Ni (VNEI\,4). We let the temperatures for those four components be independent free parameters. We link the ionization parameters ($n_{\rm e}t$) for the first three components,  but fix to $10^{10} {\rm cm^{-3}\,s}$ for the Fe--Ni plasma because of the low ionization. As a result, $\chi^2_\nu$ (d.o.f.) is improved to be 1.16 (1998) (panel:(c)).  The improvement is significant with the F-test null probability much less than 0.01\,\%.  In the $\chi^2$ test, this model is still unacceptable, possibly due to systematic errors.  In fact, we obtain no significant improvement by adding further components or by changing the trial model. We thus adopt this model as the best approximation. The best-fit parameters are given in table \ref{src-para}.

\subsubsection{Spatial analysis of the SNR}

\begin{figure}
    \begin{minipage}{0.15\hsize}
    \begin{center}
    \end{center}
    \end{minipage}
	\begin{minipage}{0.35\hsize}
	\begin{center}
		\FigureFile(55mm,55mm){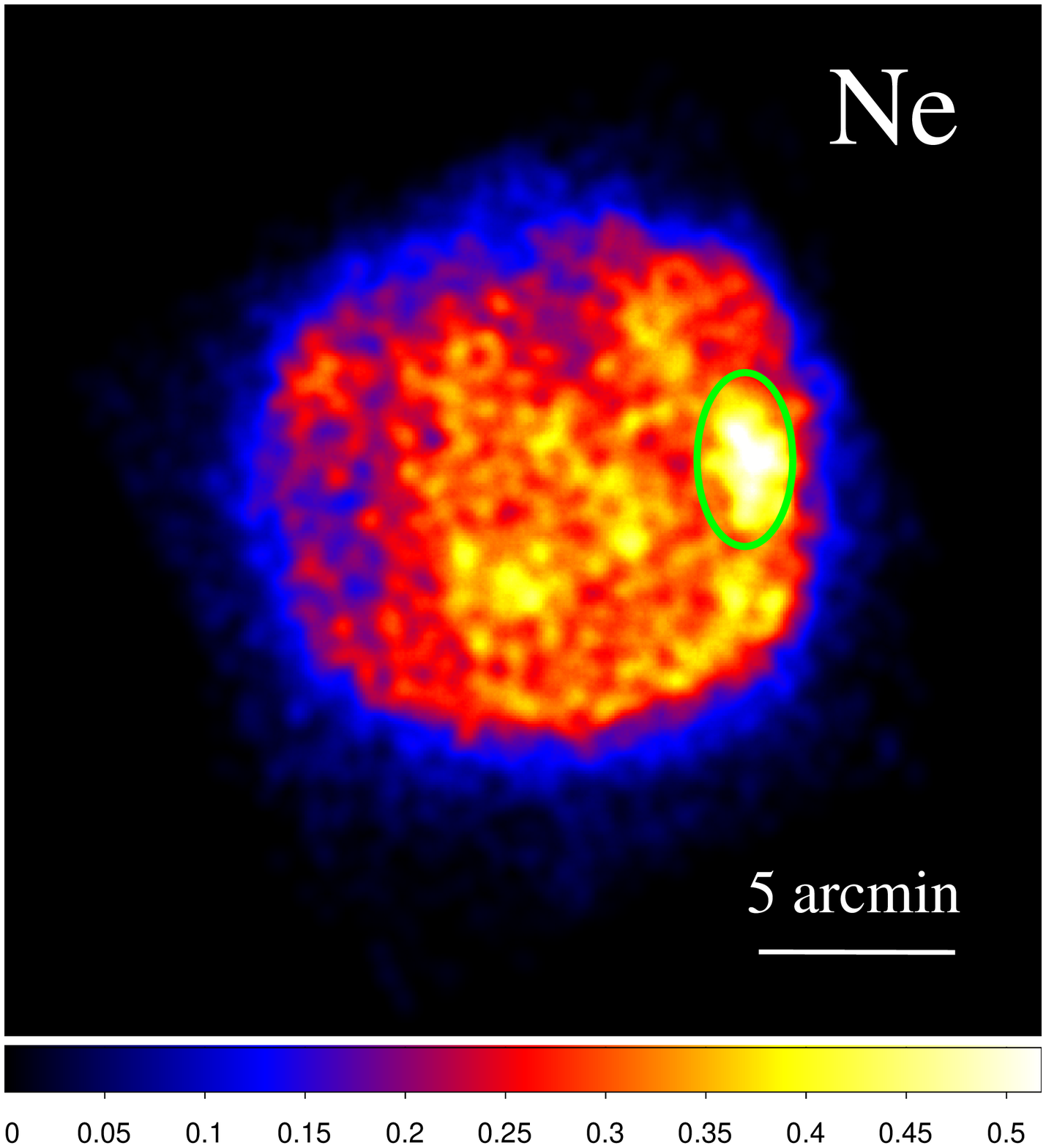}
	\end{center}
	\end{minipage}
	\begin{minipage}{0.35\hsize}
	\begin{center}
		\FigureFile(55mm,55mm){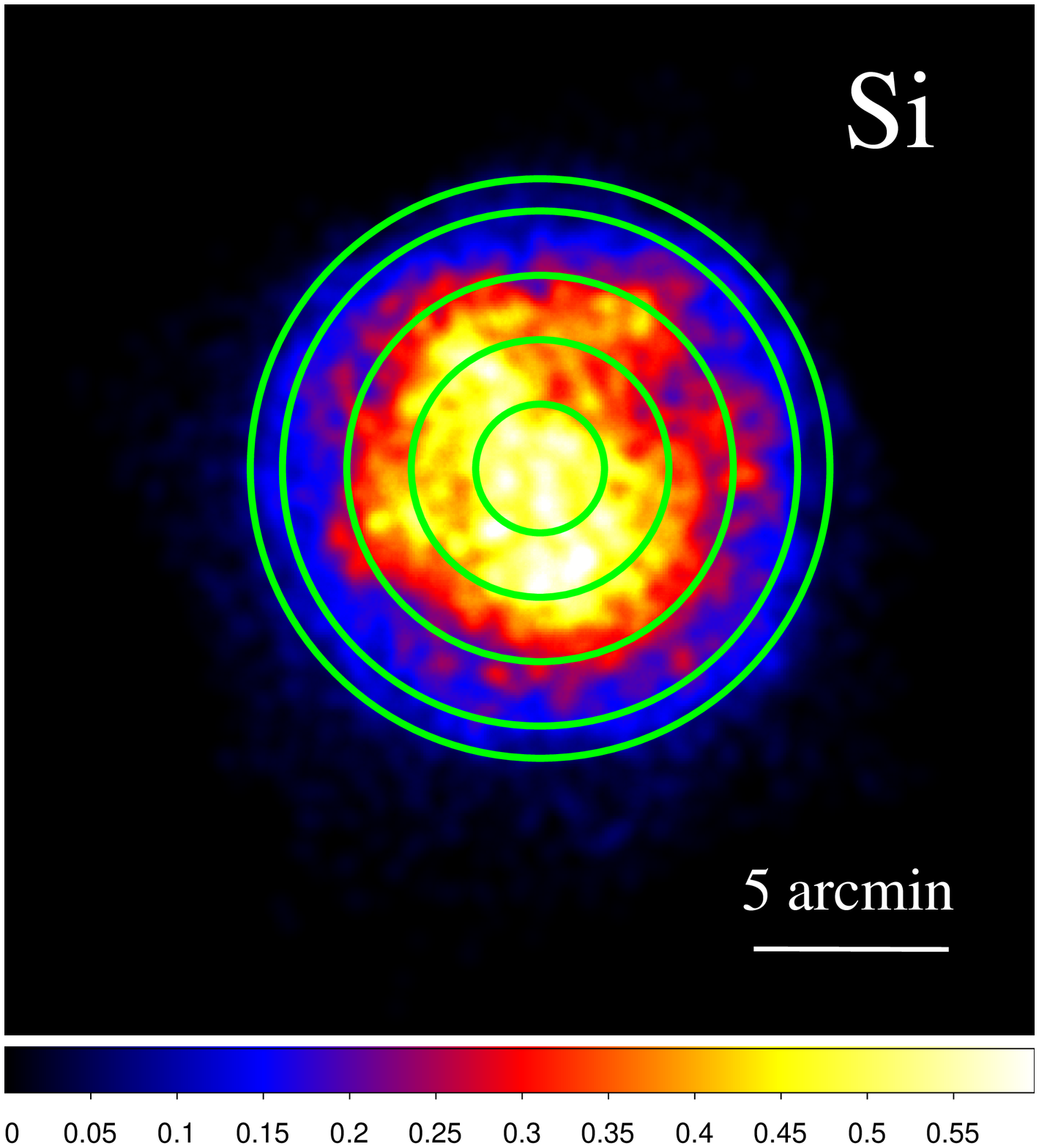}
	\end{center}
	\end{minipage}
	\caption{XIS images of G272.2$-$3.2 in the 0.86--0.96\,keV band (Ne-band: left) and 1.79--1.93\,keV band (Si-band: right), smoothed with a Gaussian kernel of $\sigma =$\,\timeform{0'.45}. The spectral extraction regions are shown with the green lines.}
\label{image}
\end{figure}

\begin{figure}
	\begin{center}
		\FigureFile(130mm,130mm){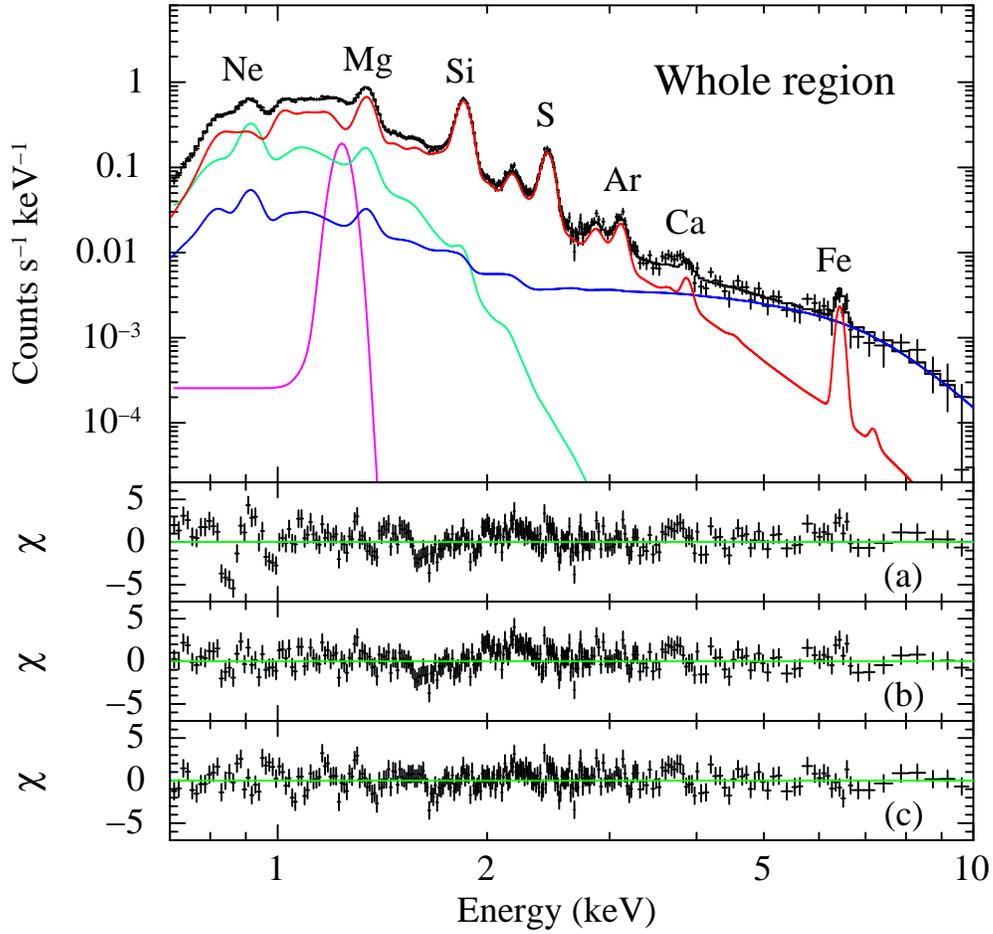}
	\end{center}
	\caption{Whole region spectrum of G272.2$-$3.2. Only the FI spectrum is displayed for the visibility. The spectrum is fitted with the model of 4-VNEI (red) + 1-APEC (green), adding the XBG (blue) and a Gaussian line (magenta). The lower panels represent the residuals from the source models of (a) 1-VNEI, (b) 1-VNEI + 1-APEC, and (c) 4-VNEI + 1-APEC, respectively.}
\label{whole_spec}
\end{figure}

\begin{figure}
	\begin{minipage}{0.5\hsize}
	\begin{center}
		\FigureFile(80mm,80mm){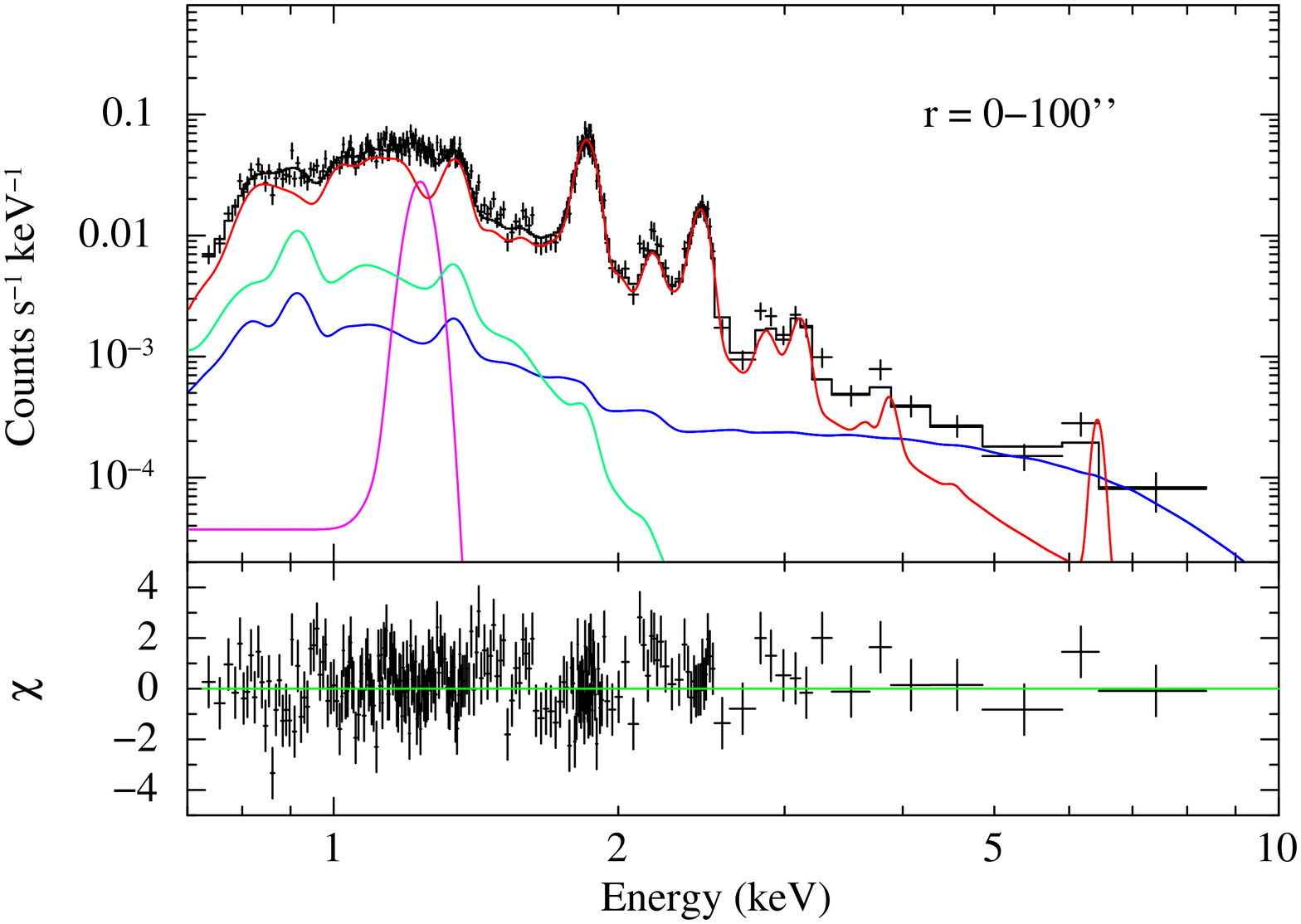}
	\end{center}
	\end{minipage}
	\begin{minipage}{0.5\hsize}
	\begin{center}
		\FigureFile(80mm,80mm){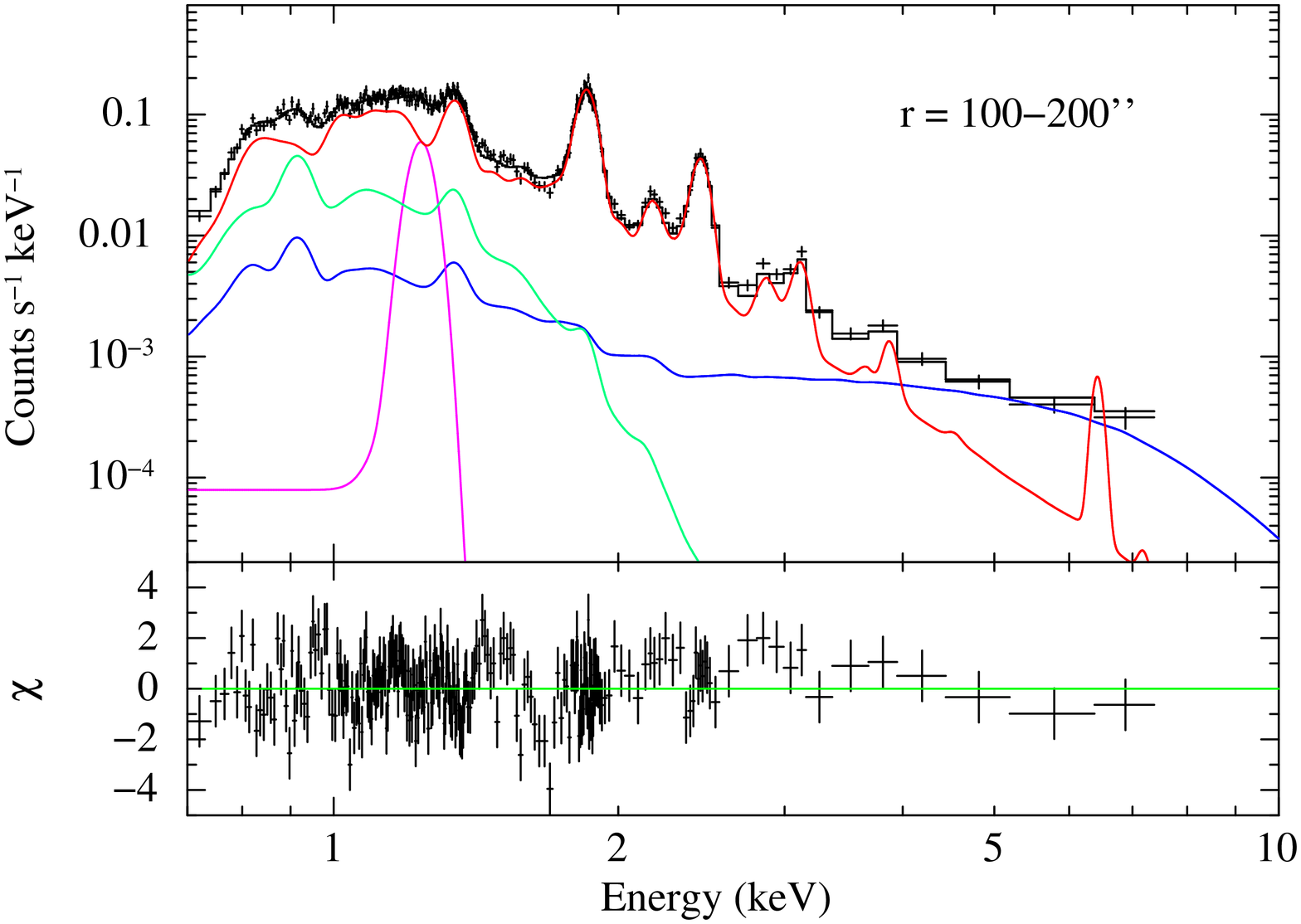}
	\end{center}
	\end{minipage}
	\begin{minipage}{0.5\hsize}
	\begin{center}
		\FigureFile(80mm,80mm){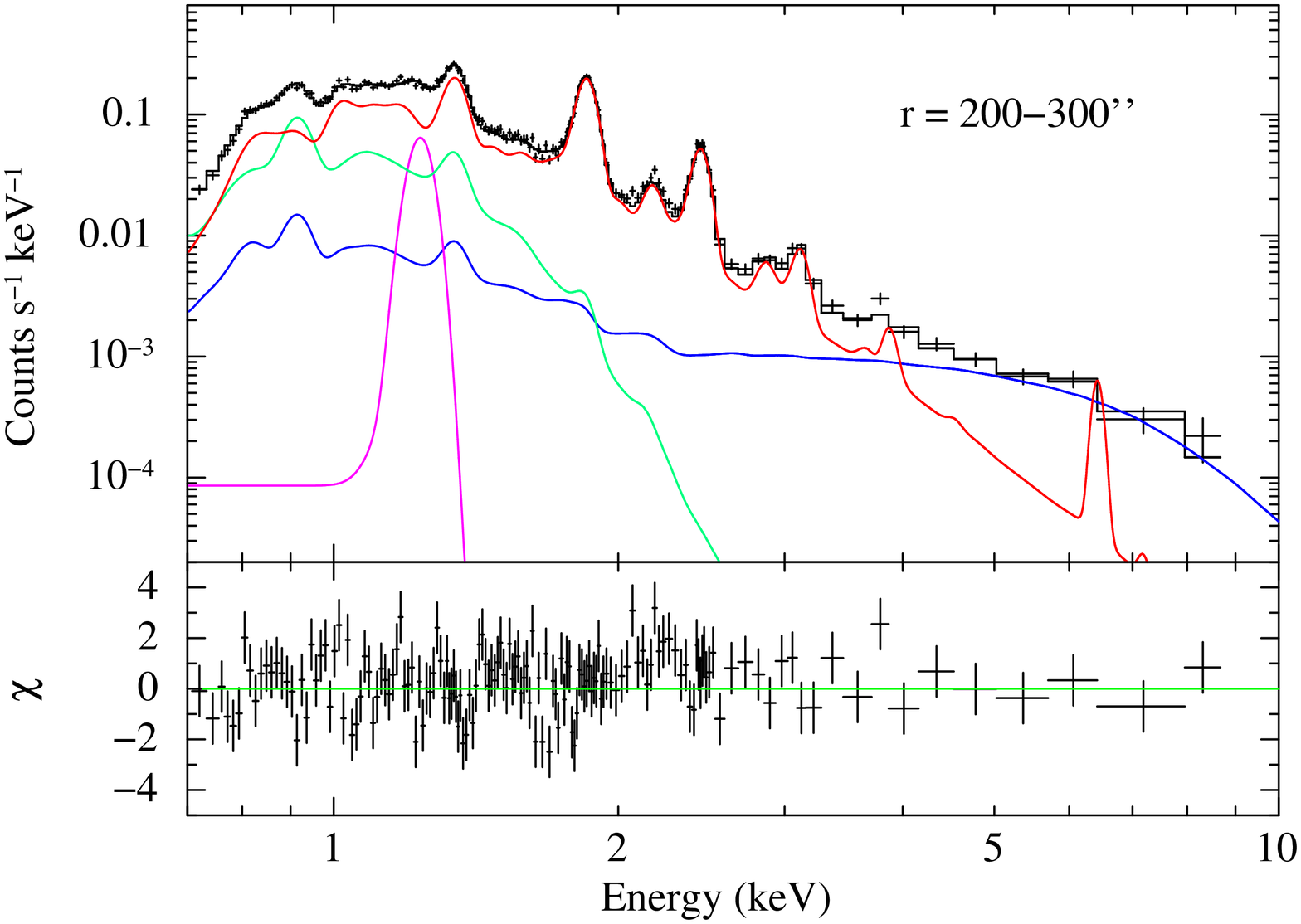}
	\end{center}
	\end{minipage}
	\begin{minipage}{0.5\hsize}
	\begin{center}
		\FigureFile(80mm,80mm){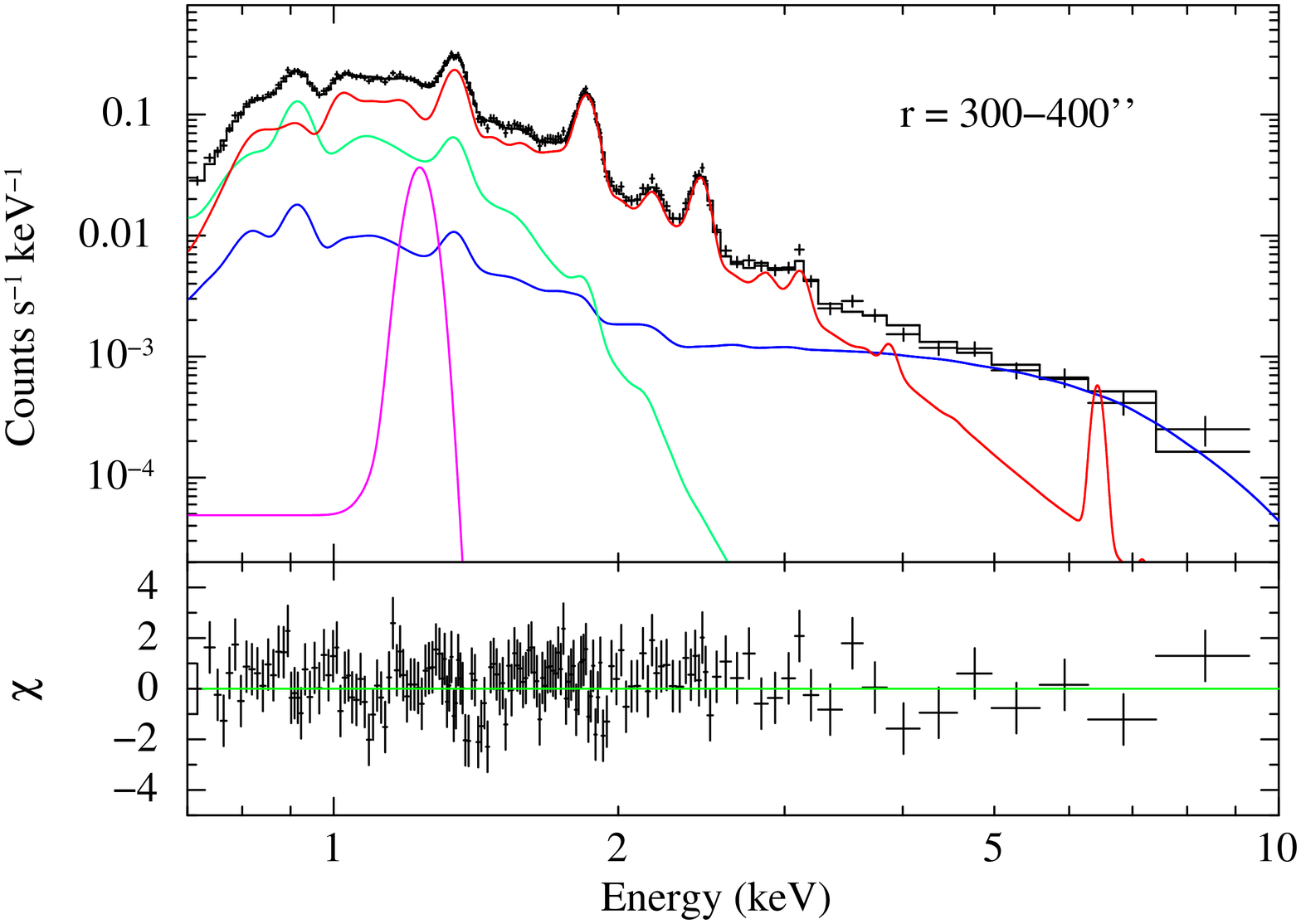}
	\end{center}
	\end{minipage}
	\begin{minipage}{0.5\hsize}
	\begin{center}
		\FigureFile(80mm,80mm){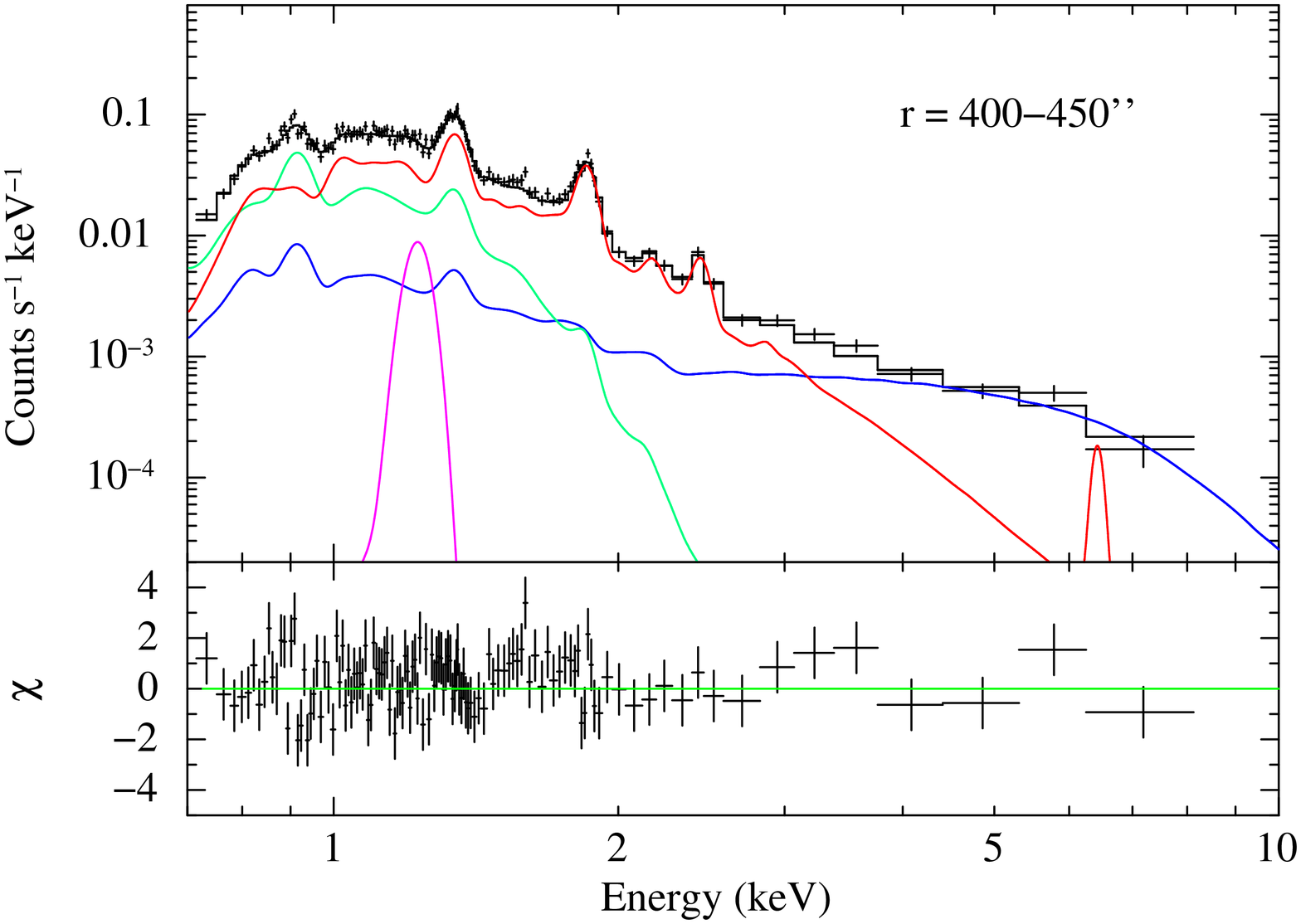}
	\end{center}
	\end{minipage}
		\begin{minipage}{0.5\hsize}
	\begin{center}
		\FigureFile(80mm,80mm){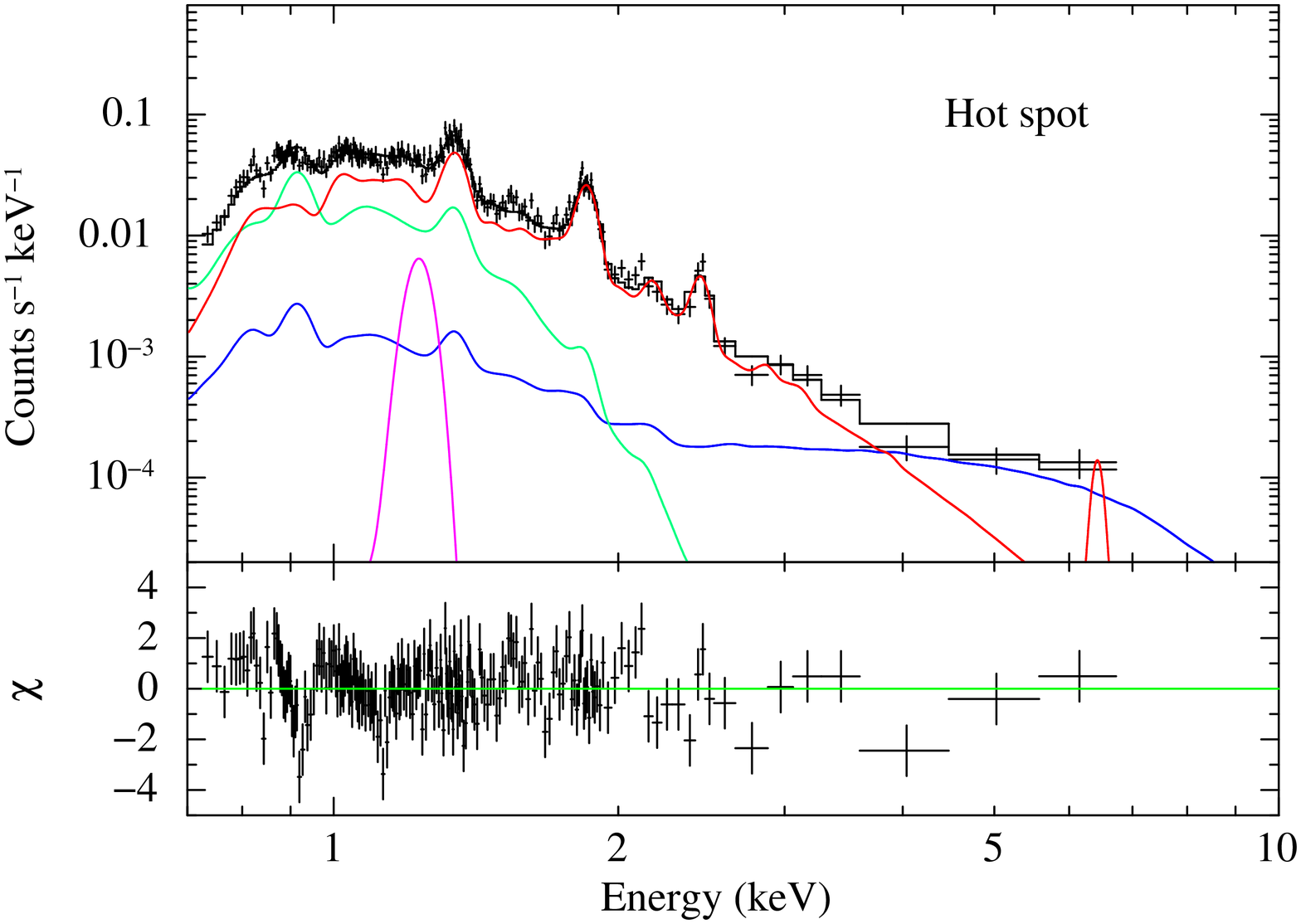}
	\end{center}
	\end{minipage}
	\caption{Spectra obtained from six regions: $r = \timeform{0-100''}$, $\timeform{100-200''}$, $\timeform{200-300''}$, $\timeform{300-400''}$, $\timeform{400-450''}$, and hot spot. Only the FI spectra are displayed for the visibility. The same model as in the figure \ref{whole_spec} is applied. Each lower panel shows the residual of the data from the model. }
\label{spectra}
\end{figure}

We next make spatially-resolved spectra dividing the SNR into a circle and four rings with the boundary radii of $r = \timeform{0-100''}$, $\timeform{100-200''}$, $\timeform{200-300''}$, $\timeform{300-400''}$ and $\timeform{400-450''}$ (figure \ref{image} right). We also define the region for the hot spot at the western edge of the remnant (figure \ref{image} left).  The spectra extracted after the NXB subtraction are shown in figure \ref{spectra}. We fit these spectra with the same model and method as those for the whole region, but fixing the column density, temperatures, and ionization parameters to the best-fit values of the whole region. We also fix the abundance of the APEC to 1 solar because of the large statistical error. The best-fit parameters are given in table \ref{src-para}.

\begin{figure}
   \begin{center}
         \FigureFile(100mm,100mm){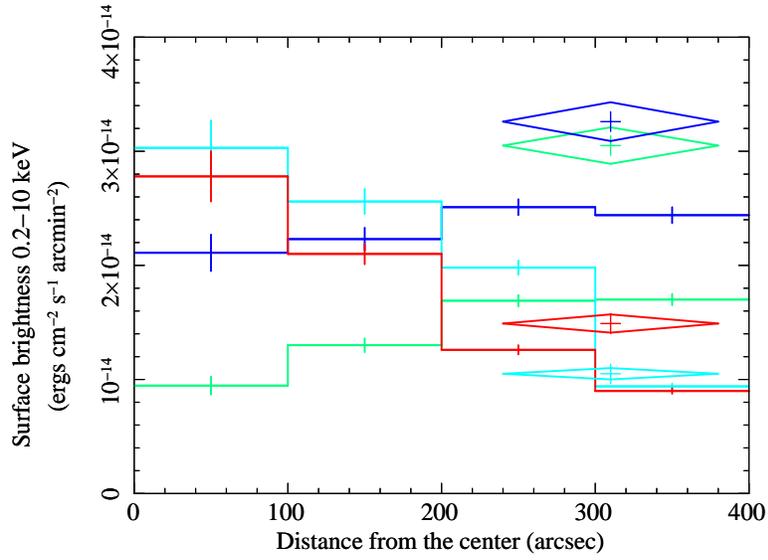}
   \end{center}
   \caption{The radial profiles of the surface brightness of the plasma components: ISM (green), H--Mg ejecta (blue), Si--Ca ejecta (light blue), and Fe--Ni ejecta (red). 
The results of the hot spot are added in diamonds.}
\label{surface_brightness} 
\end{figure}

We show the radial profile of the surface brightness of the APEC and the VNEI\,1--4 (H--Mg, Si--Ca, Fe--Ni) in figure \ref{surface_brightness}. The data from the hot spot are also added in diamonds, while the data from the  $\timeform{400-450''}$ ring are not plotted, because this region is just at the boundary of the SNR. The components show different spatial structures with each other; the surface brightness of the APEC and light elements (H--Mg) in the VNEI increase to the outer region, while those of the intermediate (Si--Ca) and Fe-peaked elements (Fe--Ni) in the VNEI concentrate at the inner region. The hot spot shows clear enhancement in the APEC and light elements in the VNEI, while it shows no clear enhancement in the intermediate and Fe-peaked elements in the VNEI.

\begin{figure}
   \begin{center}
         \FigureFile(100mm,100mm){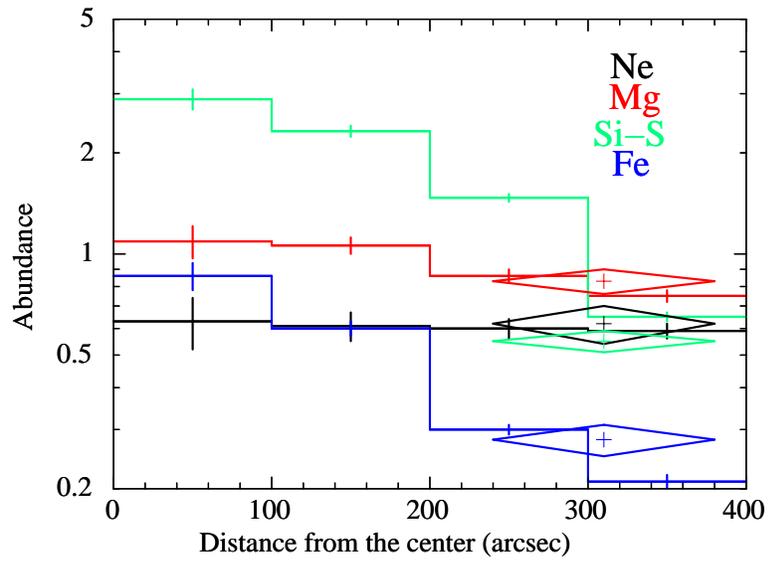}
   \end{center}
   \caption{Radial distribution of the ejecta abundances. The abundances are taken from VNEI\,1-4 plasma. The results of the hot spot are shown by the diamonds.}
\label{abund_dist} 
\end{figure}

Figure \ref{abund_dist} shows the radial distribution of the abundances of Ne, Mg, Si--S and Fe in the VNEI. The abundance of Ne is constant across the radius, while those of Si--S and Fe show central concentrations. The abundance of Mg is in between that of Ne and those of Si--S.  The abundances at the hot spot show no deviation from the radial profile.

\section{Discussion} 

We find two important facts on the Galactic SNR G272.2$-$3.2 by the revised data reduction and background subtraction for the Suzaku data.
(1) A clear emission above 3\,keV is detected, and the K-shell lines of Ar, Ca, and Fe are discovered for the first time. (2) The G272.2$-$3.2 plasma consists of two components: a low-temperature CIE plasma ($kT_{\rm e} \sim 0.2\,{\rm keV}$) and a high-temperature NEI plasma ($kT_{\rm e} = 0.6$--$3\,{\rm keV}$). 
Since the CIE plasma (APEC) has roughly solar abundances over the entire area (table \ref{src-para}), this likely originates from the ISM. The abundances of the NEI plasma (VNEI) increase toward the center and becomes higher than solar values. Therefore it must be ejecta origin.  The newly detected K-shell lines of Ar, Ca, and Fe mainly come from the ejecta.  In the following subsections, we separately discuss the ISM and the ejecta based on the revised results of the distance, mass and morphology.

\subsection{Distance estimation} 

The distance to G272.2$-$3.2 from the Sun has been estimated as 2-10\,kpc using the statistical analysis (lower limit) and the interstellar absorption (upper limit) (\cite{Harrus2001a}). The best-fit $N_{\rm H}$ of $1.0\times10^{22}\,{\rm cm^{-2}}$ is consistent with \citet{Harrus2001a}, and hence the same method and estimation of the distance (10\,kpc) may be applied. However, the observed $N_{\rm H}$ of G272.2$-$3.2 is larger than those of the sources on the plane in $l = 260$--$290^\circ$ that are summarized in table \ref{distance}, in spite of its large Galactic latitude. In fact, Pup A (G260.4$-$3.4), at similar Galactic latitude to G272.2$-$3.2, has a small value of $N_{\rm H}$ ($ <10^{22}\,{\rm cm^{-2}}$). The large value of $N_{\rm H}$ of G272.2$-$3.2 would be due to a local absorption in the line of sight at the ``Vela region" (\cite{Lallement2014a}).  On the other hand, more than 90\,$\%$ of the Galactic SNRs are statistically distributed within 100\,pc from the Galactic plane (\cite{Ilovaisky1972a}). The large Galactic latitude limits the distance of G272.2$-$3.2 to be $<2$\,kpc same as that in the statistical analysis by \citet{Harrus2001a}. Additionally, all the sources in table \ref{distance} may locate on the Carina nebula arm, like pulsars (in figure 1 of \cite{Taylor1993b}).  G272.2$-$3.2 may also locate on the edge of this arm, at the tangential point of the arm with the distance of $\sim $2--3\,kpc. Thus we assume that the most likely distance, $d$, is 2--3 kpc. Hereafter we adopt the distance of 2.5 kpc, or parameterized as $d_{2.5} = d/2.5\,{\rm kpc}$.

\subsection{ISM structure and density} 

The surface brightness of the ISM shows an increase toward the outer rings. From the angular size, the radius of G272.2$-$3.2 is $1.7\times10^{19}\,{\rm cm}$ and volume of the plasma is $2\times10^{58}\,{\rm cm^3}$.  The best-fit ISM temperature of $kT_{\rm e} = 0.17$\,keV gives the expansion speed of $V_s = 3.8\times10^2\,{\rm km\,s^{-1}}$ from the strong shock relation assuming electron-ion temperature equilibration, $kT_{\rm e} = 3/16 \mu m_{\rm H} V_s^2$, where $\mu$ and $m_{\rm H}$ are the mean atomic mass and the mass of a hydrogen atom, respectively. Consequently, the dynamical age is $\sim 6\times 10^3$\,years from the Sedov self-similar solution (\cite{Sedov1959a}).  We calculate the density of the pre-heated ISM to be $1.0\,{\rm cm^{-3}}$ from the emission measure.  This high ambient density and the large latitude of $-3.2^\circ$ imply that  G272.2$-$3.2 is located in the Carina nebula arm, consistent with the small distance of 2.5 kpc estimated in section 4.1.

\subsection{Fe K-shell line} 

The most important result of this work is the discovery of the Fe K-shell line. The Fe K-shell line energy is $6.42_{-0.05}^{+0.07}\,{\rm keV}$, and hence Fe is in a low-ionization state. The best-fit temperature ($kT_{\rm e}$) is  $2.76\pm0.15\,{\rm keV}$ with a small ionization parameter ($n_{\rm e}t$) of $10^{10}$\,cm$^{-3}$\,s. Therefore the Fe ejecta should be recently heated by the reverse shock. Recently, \citet{Yamaguchi2014a} suggested the new observational diagnostic to discriminate the progenitor types of SNRs using the Fe K-shell lines. They argued that the Fe K-shell lines of CC SNRs have higher center energies (6.6--6.7\,keV) than those of Type Ia SNRs (6.4--6.5\,keV). According to their criterion, G272.2$-$3.2 is a member of Type Ia SNRs.

\subsection{Radial profile of the ejecta} 

The radial distribution of the surface brightness of Si--Ca and Fe in the ejecta increases toward the center. The abundances of the ejecta also have a layer-like structure (figure \ref{abund_dist});  the radial pattern of Ne is rather flat compared to the central concentration of Si--S and Fe. The best-fit temperatures in the NEI plasma of the element group (H--Mg, Si--S, Ar--Ca, Fe--Ni), on the other hand, show smooth increase in the order of the atomic mass (table 2):  $kT_{\rm e} = 0.62\pm0.02$\,keV (H--Mg), $0.80\pm0.03$\,keV (Si--S), $1.00\pm0.22$\,keV (Ar--Ca), and $2.76\pm0.15$\,keV (Fe--Ni). These results suggest that the element groups are stratified. The stratified structure of the elements and the temperature distribution of the relevant plasma are consistent with numerical simulations of Type Ia SNe (e.g., \cite{Dwarkadas1998a}). We note that \citet{Lopez2011a} argued a Type Ia origin of G272.2$-$3.2 by the spherically symmetric structure, in contrast to the fact that CC SNRs generally have deformed shapes due to the interaction with the complex and dense environments, where the massive progenitors are born. 

\subsection{Ejecta mass} 

If the ejecta are really a Type Ia origin, the plasma has no H nor He. With this assumption, we re-fit the whole region spectrum. The $\chi^2_\nu$ (d.o.f.) becomes to be 1.16~(1999), that is equal to that of H-dominant plasma (1.16~(1998)). The best-fit parameters of the ejecta (VNEI) are almost the same as those of table \ref{src-para} (Whole),  but the normalization becomes to be $\int n_e n_{\rm C} dV / (4 \pi d^2) = 2.8\times10^{9}\,{\rm cm^{-5}}$, where $n_{\rm C}$ represents the density of carbon. 
Using the SPEX software (\cite{Kaastra1996a}), we calculate $n_{\rm e}$ and $n_{\rm C}$ as $5\times10^{-2}\,(d_{2.5})^{-0.5}\,f^{-0.5}\,{\rm cm}^{-3}$ and $2\times10^{-3}\,(d_{2.5})^{-0.5}\,f^{-0.5}\,{\rm cm}^{-3}$, respectively, where $f$ is a filling factor. Therefore, the mass of the non-Fe ejecta is estimated to be $\sim1.8\,(d_{2.5})^{2.5}\,f^{0.5}\,M_\odot$, while that of the Fe ejecta is only $\sim 6 \times~10^{-2}\,(d_{2.5})^{2.5}\,f^{0.5}\,M_\odot$.  The mass of the non-Fe ejecta is two or three times larger than that expected in a Type Ia SN (0.5--0.9\,$M_\odot$; e.g., \cite{Maeda2010a}). However, adopting the lower limit of the distance of 2\,kpc (see section 4.1) and the filling factor of $1 / 4$, the non-Fe mass becomes to be $\sim0.5\,M_\odot$ that is consistent with the Type Ia model. The mass of Fe is, on the other hand, too small, compared with that of the Type Ia. It may be due to the fact that only a small fraction of Fe in the center regions is heated by the reverse shock, which is often seen for young Type Ia SNRs (e.g., N103B: \cite{Lewis2003a}; SN\,1006: \cite{Yamaguchi2008a}).

\subsection{Hot spot} 

G272.2$-$3.2 shows a spherical symmetric shape, the typical morphology of Type Ia SNRs (\cite{Lopez2011a}).  We, however, find one singularity, which is the bright hot spot at the western edge of this SNR.  From table \ref{src-para}, the surface brightness of the hot spot shows significant enhancement in the ISM plasma and Ne--Mg in the ejecta plasma. On the other hand, the abundances show no large deviation from the mean radial profile (figure \ref{abund_dist}).  Thus the origin of the hot spot would be a dense ISM. It must be heated up by the light elements of ejecta located in the outer layer.

 
\section{Summary}

We have analyzed the Suzaku/XIS data of the Galactic SNR, G272.2$-$3.2. The results are summarized as follows:

\begin{enumerate}

\item   The X-ray emission from G272.2$-$3.2 consists of two types of components, one is a CIE plasma and the other is a NEI plasma.

\item  The CIE plasma has roughly solar abundances, and hence is likely the ISM origin, while the NEI plasma has super-solar in the central region, indicating the ejecta origin.

\item  In the NEI plasma, we discover the K-shell lines of Ar, Ca and Fe. The Fe is in a low ionization state, which would be recently heated by a reverse shock.

\item  The abundances of the ejecta have different radial distributions; Ne is almost constant across the radius, while Si--Ca and Fe concentrate at the center. The lighter elements have lower temperatures than those of the heavier elements.
 
\item  Based on the morphologies of the ejecta and ISM, the presence of nearly neutral Fe and the ejecta mass, we conclude that the origin of G272.2$-$3.2 is a Type Ia SN rather than a CC SN.

\end{enumerate}

\section*{Acknowledgment}

We thank all members of the Suzaku operation and calibration teams. This work is supported by Japan Society for the Promotion of Science (JSPS) KAKENHI Grant Number 15H03641, 15J015760, 23000004, 23340071, 24540229, 24684010,  24740167, 25800119, 26670560, 26800102. 

\newpage
\bibliographystyle{pasj}
\bibliography{reference}

\newpage

\newpage

\begin{table*}
\caption{Best-fit parameters for the background spectrum.}\label{bkg-para}
\begin{center}
\begin{tabular}{llc}
\hline\hline
Component & \multicolumn{1}{c}{Parameter} & Value\\
\hline
\multicolumn{3}{c}{$Abs1 \times CXB + Abs2 \times GH ^*$}\\ \hline
Abs1 & $N_{\rm H}$ $(10^{21} {\rm cm^{-2}})$ & 8.6 (fixed) \\
CXB & Photon index & 1.4 (fixed) \\
        & Flux$^\dag$ & $2.44\pm0.09$ \\
Abs2 & $N_{\rm H}$ $(10^{21} {\rm cm^{-2}})$ & $8.0\pm0.5$ \\
GH$^\ddag$ & $kT_{\rm e}$ (keV) & $0.20\pm0.02$\\
\hline
$\chi^2_\nu$~(d.o.f.) & & 1.19 (360)\\
\hline
\multicolumn{3}{l}{\small $^*$ Abbreviations representation of the background model.}\\[-1.5mm]
\multicolumn{3}{l}{\small Each component represents the interstellar absorption (Abs),}\\[-1.5mm]
\multicolumn{3}{l}{\small the cosmic X-ray background (CXB), and the Galactic halo (GH).}\\[-1.5mm]
\multicolumn{3}{l}{\small $^\dag$ Flux ($10^{-11}\,{\rm erg\,cm^{-2}\,s^{-1}\,deg^{-2}}$) in the $2-10$\,keV band.}\\[-1.5mm]
\multicolumn{3}{l}{\small $^\ddag$ Abundances are fixed to the solar value}\\[-1.5mm]
\multicolumn{3}{l}{\small (\cite{Anders1989a}).}
\end{tabular}	
\end{center}
\end{table*}

\newpage

\newpage

\begin{table*}
\caption{Best-fit parameters for the SNR spectra.}\label{src-para}
\begin{center}
\footnotesize
\begin{tabular}{lcccccccc}
\hline\hline
Component & Parameter & Whole & 0--\timeform{100''} & 100--\timeform{200''} & 200--\timeform{300''} & 300--\timeform{400''} & 400--\timeform{450''} & Hot-spot\\
\hline
Absorption & $N_{\rm H}$ $(10^{22} {\rm cm^{-2}})_{\rm FI}$ & $0.99\pm0.01$ & \multicolumn{6}{c}{0.99 (fixed)}\\
  & $N_{\rm H}$ $(10^{22} {\rm cm^{-2}})_{\rm BI}$ & $0.98\pm0.01$ & \multicolumn{6}{c}{0.98 (fixed)}\\
VNEI\,1 & $kT_{\rm e}$ (keV) & $0.62\pm0.02$ & \multicolumn{6}{c}{0.62 (fixed)}\\
  & H & 1 (fixed) & \multicolumn{6}{c}{1 (fixed)}\\
  & He & 1 (fixed) & \multicolumn{6}{c}{1 (fixed)}\\
  & C & 1 (fixed) & \multicolumn{6}{c}{1 (fixed)}\\
  & O & 1 (fixed) & \multicolumn{6}{c}{1 (fixed)}\\
  & Ne & $0.54\pm0.03$ & $0.63\pm0.11$ & $0.61\pm0.06$ & $0.60\pm0.04$ & $0.59\pm0.03$ & $0.47\pm0.05$ & $0.62\pm0.08$\\
  & Mg & $0.74\pm0.03$ & $1.09\pm0.12$ & $1.06\pm0.06$ & $0.86\pm0.04$ & $0.75\pm0.03$ & $0.67\pm0.05$ & $0.83\pm0.07$\\
  & $n_{\rm e} t$ $(10^{10} {\rm cm^{-3}\,s})$ & $12.3\pm1.0$ & \multicolumn{6}{c}{12.3 (fixed)}\\
  & $EM^\dag$ $(10^{11} {\rm cm^{-5}})$ & $30.1\pm2.5$ & $1.28\pm0.10$ & $4.12\pm0.18$ & $8.23\pm0.24$ & $11.1\pm0.3$ & $2.75\pm0.11$ & $2.12\pm0.11$\\
VNEI\,2 & $kT_{\rm e}$ (keV) & $0.80\pm0.03$ & \multicolumn{6}{c}{0.80 (fixed)}\\
  & Si & $0.97\pm0.06$ & $2.44\pm0.22$ & $1.94\pm0.10$ & $1.23\pm0.04$ & $0.63\pm0.02$ & $0.50\pm0.03$ & $0.60\pm0.05$\\
  & S & $1.25\pm0.08$ & $3.34\pm0.33$ & $2.69\pm0.15$ & $1.70\pm0.07$ & $0.66\pm0.03$ & $0.40\pm0.05$ & $0.50\pm0.07$\\
  & $n_{\rm e} t$ $(10^{10} {\rm cm^{-3}\,s})$ & & \multicolumn{6}{c}{(Linked to VNEI\,1)}\\
  & $EM^\dag$ $(10^{11} {\rm cm^{-5}})$ & & \multicolumn{6}{c}{(Linked to VNEI\,1)}\\
VNEI\,3 & $kT_{\rm e}$ (keV) & $1.00\pm0.22$ & \multicolumn{6}{c}{1.00 (fixed)}\\
  & Ar = Ca & $0.68\pm0.22$ & $1.73\pm0.34$ & $1.58\pm0.17$ & $1.02\pm0.10$ & $0.38\pm0.07$ & $< 0.13$ & $< 0.18$\\
  & $n_{\rm e} t$ $(10^{10} {\rm cm^{-3}\,s})$ & & \multicolumn{6}{c}{(Linked to VNEI\,1)}\\
  & $EM^\dag$ $(10^{11} {\rm cm^{-5}})$ & & \multicolumn{6}{c}{(Linked to VNEI\,1)}\\
VNEI\,4 & $kT_{\rm e}$ (keV) & $2.76\pm0.15$ & \multicolumn{6}{c}{2.76 (fixed)}\\
  & Fe = Ni & $0.30\pm0.02$ & $0.86\pm0.08$ & $0.60\pm0.03$ & $0.30\pm0.01$ & $0.21\pm0.01$ & $0.22\pm0.02$ & $0.28\pm0.03$\\
  & $n_{\rm e} t$ $(10^{10} {\rm cm^{-3}\,s})$ & 1 (fixed) & \multicolumn{6}{c}{1 (fixed)}\\
  & $EM^\dag$ $(10^{11} {\rm cm^{-5}})$ & & \multicolumn{6}{c}{(Linked to VNEI\,1)}\\
APEC & $kT_{\rm e}$ (keV) & $0.172\pm0.005$ & \multicolumn{6}{c}{0.172 (fixed)}\\
  & All elements & $< 2.4$ & \multicolumn{6}{c}{1 (fixed)}\\
  & $EM^\dag$ $(10^{11} {\rm cm^{-5}})$ & $342\pm33$ & $11.4\pm1.7$ & $46.9\pm2.8$ & $101.3\pm3.6$ & $137.0\pm4.0$ & $39.1\pm1.8$ & $36.1\pm2.0$\\
\hline
$\chi^2_\nu$~(d.o.f.) & & 1.16~(1998) & 1.09~(712) & 1.15~(1186) & 1.14~(1392) & 1.04~(1366) & 1.03~(870) & 1.13~(620)\\
\hline
\multicolumn{8}{l}{\small $^\dag$ Emission measure defined as $\int n_e n_{\rm H} dV / (4 \pi d^2)$, where {\it{V}} and {\it{d}} are the emitting volume (cm$^3$) and the distance to}\\[-1.5mm]
\multicolumn{8}{l}{\small the source (cm), respectively.}\\[-1.5mm]
\end{tabular}
\end{center}
\end{table*}

\begin{table*}
\caption{Column densities of the sources on the plane of $l = 260-290$\timeform{D}.}\label{distance}
\begin{center}
\footnotesize
\begin{tabular}{lcccc}
\hline\hline
Name & ($l$, $b$) &$d$ & $N_{\rm H}$ & Refs.$^*$\\[-1.5mm]
  & & (kpc) & ($10^{21}\,{\rm cm^{-2}}$) & \\
\hline
Puppis A & ($\timeform{260D.4}$, $\timeform{-3D.4}$) & 2.2 & 2 -- 4 & 1, 2, 3, 4\\
Vela SNR & ($\timeform{263D.9}$, $\timeform{-3D.3}$) & 0.25 -- 0.3 & 0.2 -- 0.4 & 5, 6\\
Vela Jr  & ($\timeform{266D.2}$, $\timeform{-1D.2}$) & 0.5 -- 1 & 3 -- 6 & 7, 8, 9, 10\\
G272.2$-$3.2  & ($\timeform{272D.2}$, $\timeform{-3D.2}$) & 2 -- 10 & 9 -- 11 & 11, 12, 13, 14\\
PSR J1016-5857  & ($\timeform{284D.0}$, $\timeform{-1D.8}$) & 3 & 12 & 15, 16\\
MSH 10-53  & ($\timeform{284D.3}$, $\timeform{-1D.8}$) & 2.9 & 8 & 17, 18\\
XMMUJ101855.4 & ($\timeform{284D.3}$, $\timeform{-1D.8}$) & 5.4 & 6.6 & 18\\[-1.5mm]
-58564 \\
B1046$-$58  & ($\timeform{287D.4}$, $\timeform{+0D.6}$) & 3 & 4 -- 9 & 16, 19, 20\\
MSH 11-61A  & ($\timeform{290D.1}$, $\timeform{-0D.8}$) & 7 & 5 -- 9 & 21, 22, 23\\
MSH 11-62  & ($\timeform{291D.0}$, $\timeform{-0D.1}$) & 1 -- 11 & 6 -- 6.7 & 24, 25\\
MSH 11-54  & ($\timeform{292D.0}$, $\timeform{+1D.8}$) & 6 & 4 -- 6 & 26, 27, 28, 29\\
\hline
\multicolumn{5}{l}{\small $^*$ References: 
(1) \cite{Reynoso2003a}; (2) \cite{Hwang2008a};}\\
\multicolumn{5}{l}{\small (3) \cite{Katsuda2008a}; (4) \cite{Katsuda2010a}; (5) \cite{Lu2000a};}\\
\multicolumn{5}{l}{\small (6) \cite{Dodson2003a}; (7) \cite{Katsuda2008b} (8) \cite{Pannuti2010a}; }\\
\multicolumn{5}{l}{\small (9) \cite{Kishishita2013a}; (10) \cite{Allen2015a}; (11) \cite{Greiner1994a}; }\\
\multicolumn{5}{l}{\small (12) \cite{Harrus2001a}; (13) \cite{Sanchez-Ayaso2013a}; } \\
\multicolumn{5}{l}{\small (14) \cite{McEntaffer2013a}; (15) \cite{Camilo2004a};} \\
\multicolumn{5}{l}{\small (16) \cite{Kargaltsev2008a}; (17) \cite{Ruiz1986a}; } \\
\multicolumn{5}{l}{\small (18) \cite{Abramowski2012a}; (19) \cite{Cordes2002a}; }\\
\multicolumn{5}{l}{\small (20) \cite{Gonzalez2006a}; (21) \cite{Filipovic2005a}; (22) \cite{Garcia2012a};}\\
\multicolumn{5}{l}{\small (23) \cite{Kamitsukasa2015a}; (24) \cite{Harrus1998a}; (25) \cite{Slane2012a};}\\ 
\multicolumn{5}{l}{\small (26) \cite{Gaensler2003a}; (27) \cite{Park2004a}; (28) \cite{Lee2010a}}\\
\multicolumn{5}{l}{\small (29) \cite{Kamitsukasa2014a};}\\
\end{tabular}
\end{center}
\end{table*}

\end{document}